\documentclass[structabstract]{aa}
\usepackage{graphicx}
\usepackage{txfonts}

\begin{document}

\title{On the effect of cosmic rays \\ in bolometric CMB measurements \\ from the stratosphere}

\author{S. Masi\inst{1,2}, E. Battistelli\inst{1,2}, P. de Bernardis\inst{1,2}, L.
Lamagna\inst{1,2}, \\ F. Nati\inst{1,2}, L. Nati\inst{1,2}, P.
Natoli\inst{3}, G. Polenta\inst{4,5}, A. Schillaci\inst{1,2}}

\institute{ Dipartimento di Fisica, Universit\`{a} di Roma ``La
Sapienza", Roma, Italy
    \and INFN Sezione di Roma 1, Roma, Italy
    \and Dipartimento di Fisica, Universit\`{a} di ``Tor Vergata", Roma, Italy
    \and Agenzia Spaziale Italiana - ASI Science Data Center, Frascati, Italy
    \and INAF - Osservatorio Astronomico di Roma, Monte Porzio Catone, Italy }

\offprints{silvia.masi@roma1.infn.it}
\date{Submitted: Jan. 14, 2010, Accepted: May 11, 2010}

\abstract {Precision measurements of the anisotropy of the cosmic
microwave background (CMB) are able to detect low-level
non-Gaussian features caused by either topological defects or the
inflation process. These measurements are becoming feasable with
the development of large arrays of ultra-sensitive bolometric
detectors and their use in balloon-borne or satellite missions.
However, the space environment includes a population of cosmic
rays (CRs), which produce spurious spikes in bolometric signals.}
{We analyze the effect of CRs on the measurement of CMB anisotropy
maps and the estimate of cosmological non-Gaussianity and angular
power spectra of the CMB.} {Using accurate simulations of noise
and CR events in bolometric detectors, and de-spiking techniques,
we produce simulated measured maps and analyze the Gaussianity and
power spectrum of the maps for different levels and rates of CR
events.} {We find that a de-spiking technique based on outlier
removal in the detector signals contributing to the same sky pixel
is effective in removing CR events larger than the noise. However,
low level events hidden in the noise produce a positive shift of
the average power signal measured by a bolometer, and increase its
variance. If the number of hits per pixel is large enough, the
data distribution for each sky pixel is approximately Gaussian,
but the skewness and the kurtosis of the temperatures of the
pixels indicate the presence of some low-level non-Gaussianity.
The standard noise estimation pipeline produces a positive bias in
the power spectrum at high multipoles.} {In the case of a typical
balloon-borne survey, the CR-induced non-Gaussianity will be
marginally detectable in the membrane bolometer channels, but be
negligible in the spider-web bolometer channels.  In experiments
with detector sensitivity better than 100 $\mu K / \sqrt{Hz}$, in
an environment less favorable than the earth stratosphere, the
CR-induced non-Gaussianity is likely to significantly affect the
results.}


\authorrunning{Masi \emph{et al.}}
\titlerunning{Cosmic rays events in bolometric CMB observations}
\maketitle

\section{Introduction}\label{introduction}

Information about the early Universe is encoded in the primary
anisotropy of the CMB \cite{}. While the Gaussian fluctuations
expected in the adiabatic inflationary scenario beautifully fit
the available power spectrum data (see \emph{e.g.} Nolta et al.
\cite{Nolt09} and Komatsu et al. \cite{Koma10}), non-Gaussian
signals are also expected at a lower level in the maps, due to
non-linearities in the inflation potential (see \emph{e.g.} Verde
et al. \cite{Verd00}) and/or the presence of topological defects
(see \emph{e.g.} Kaiser \& Stebbins \cite{Kays84}). In addition,
non-Gaussian secondary anisotropies are imprinted in the
post-recombination universe because of the interaction of CMB
photons with the large-scale structures present in the Universe,
and the emission of Galactic and extragalactic sources. The study
of primordial non-Gaussianity can in principle allow to confirm
and select an inflation model, but the signal to be detected is
very small. For this reason, it is essential to ensure that the
measurement system does not introduce non-Gaussian features into
the data.

Arrays of microwave detectors feature high mapping speed and are
now starting to operate at the focus of large telescopes (see
\emph{e.g.} Sayers et al. \cite{Saye09}, Siringo et al.
\cite{Siri09}, Carlstrom et al. \cite{Carl09}, Wilson et al.
\cite{Wils08}, Swetz et al. \cite{Swet08}) or in CMB polarimeters
(see \emph{e.g.} Hinderks et al. \cite{Hind09}, Yoon et al.
\cite{Yoon06}, Kuo et al. \cite{Kuo06}, Samtleben et al.,
\cite{Samt07}). Transition edge superconducting (TES) bolometric
detectors, involving only litographic fabrication techniques, are
easier to replicate in large arrays, and extremely sensitive.

To fully exploit their sensitivity, the radiative background has
to be minimized. In this sense, the optimal solution is to use
bolometric detectors in space, possibly feeding them by means of a
cryogenic telescope or optical system. The bolometers of the HFI
instrument on the Planck mission (Holmes et al. \cite{Holm08}) and
those of the PACS and SPIRE instruments on the Herschel mission
(Billot et al. \cite{Bill09}, Schultz et al. \cite{Schu08})
represent a first step in this direction.

Being extremely sensitive to any form of energy deposited on their
absorber and thermistors, bolometers are also sensitive to cosmic
rays. The energy deposited by a single MeV proton in a cryogenic
bolometer is much higher than the typical level of the noise (see
\emph{e.g.} Caserta et al. \cite{Case90}): if an amount of energy
$\Delta E$ is deposited in a bolometer in a time much shorter than
the bolometer time constant, the temperature rise of the bolometer
will be $\Delta T = \Delta E / C$, where $C$ is the heat capacity
of the detector. This peak level will be achieved sooner than one
time constant; the subsequent decay, instead, will be regulated by
the time constant. The typical noise of a bolometer corresponds to
physical temperature fluctuations much lower than $\Delta E / C$,
so most CR-induced spikes will be evident in the bolometer
time-ordered-data.

To allow the operation of bolometers in unprotected environments
(such as sub-orbital or orbital space instruments), special low
cross-section bolometers have been developed (spider-webs:
Mauskopf et al. \cite{mausk97}, wire-grids: Jones et al.
\cite{jones03}). An additional benefit of these low-cross-section
detectors is the reduced heat capacity with respect to
solid-absorber or membrane-absorber ones, resulting in a shorter
time constant.

In a polar stratospheric balloon flight, the rate of cosmic-ray
events in low cross-section (spider-web) bolometers cooled to 0.3K
is on the order of 0.1 Hz (Masi et al. \cite{Masi06}), which is
about 20 times less than the rate of events for
solid/membrane-absorber bolometers at the same temperature in the
same environment. The events are produced either by primary CR
interacting with the detectors, or by secondary particles,
including showers of electrons and bremmstrahlung-produced gammas,
resulting from the interactions of primary cosmic rays with the
metal surrounding the bolometric detector. Lower temperature
detectors of basically the same geometry are affected by lower
noise, such that the rate of CR events above the noise level is
even higher; this effect is in part mitigated by their more rapid
response. Rates between 0.02 Hz and 0.3 Hz have been measured for
spider-web bolometers cooled to 0.1K in similar stratospheric
conditions (see Macias-Perez et al. \cite{Maci07}). The promising
kinetic inductance detectors (KIDs), currently under development,
are also sensitive to CRs, at least if the substrate is solid Si
or sapphire. Anyway, their very fast response reduces
significantly the fraction of contaminated data. Moreover, their
sensitivity to CRs can be strongly reduced by depositing the KID
resonator on a membrane. Coherent radiometers, using macroscopic
thermally stable components, are much less prone to CR glitches.

Balloon-borne missions, using bolometer arrays, currently under
development, include among others the EBEX CMB polarimeter (Oxley
et al. \cite{Oxle04}), the SPIDER experiment to detect CMB
polarization at larger angular scales (Crill et al.,
\cite{Crill08}), the OLIMPO telescope, mainly devoted to high
frequency observations of the Sunyaev-Zeldovich effect (Nati et
al., \cite{Nati07}), and the BLAST/BLASTPOL telescope devoted to
far infrared cosmological and Galactic studies (Pascale et al.,
\cite{Pasc08}, Marsden et al., \cite{Mars08}) . Given the high
mapping speed of large arrays of bolometers and the efficient and
relatively cheap access to space provided by stratospheric
balloons, we expect these efforts to continue in the future. It is
thus relevant to focus on the stratospheric case.

The data from a bolometric detector can be cleaned by detecting
the spikes and flagging the corresponding section of data as
unusable for additional analysis. For filtering purposes, the
masked data can be filled with constrained realizations of noise
with the same properties as the surrounding data (see \emph{e.g.}
Masi et al. \cite{Masi06}). However, in the absence of independent
cosmic rays monitors, a number of low level events will remain
unidentified, hidden in the noise. If the bolometer is non-ideal,
and features an additional long time constant, the effect of CR
can be an increase in 1/f noise. In this paper, we focus on the
undetected cosmic rays hits and study how they can affect the
analysis of data aimed at measuring CMB anisotropy and
non-Gaussianity.

\section{Simulation of bolometer time-ordered data}

To study the effect of CR events, we simulated first time-streams
of bolometer data in the absence of sky signals, \emph{i.e.}, we
added CR events to a Gaussian realization of noise $G(t)$ with a
given standard deviation $\sigma$ and null average. The
time-ordered samples of bolometer noise are built as a Gaussian
realization, filtered with a first-order low-pass filter
simulating the thermal response of the detector.

We considered two reference cases with different bolometer time
constants and CR rates: case $R_1$ has a rate of 2 Hz for a slow,
membrane absorber bolometer at 0.3K ($\tau = 70 ms$), and case
$R_2$ is a low-cross-section bolometer (spider-web) with a rate of
0.1 Hz and a time constant of $\tau = 10 ms$.

The waiting times for CR events follow an exponential distribution
of rate $R$

\begin{equation}
P(t)=Re^{-Rt} ,  t>0.
\end{equation}

\noindent The amplitudes $A$ of the pulses induced by CRs are
assumed to follow an exponential distribution with average
$\langle A \rangle$. The exact shape of this distribution for a
given experiment depends on the spectra of primary cosmic rays in
the environment of operation, on the distribution and
characteristics of the material surrounding the bolometer, on the
shape of the sensitive element of the bolometer itself, and on the
angular distribution of the incoming cosmic rays in the restframe
of the bolometer sensor. The important parameter, however, is the
fraction of events that have amplitudes lower than say 3$\sigma$:
these are likely to remain hidden in the noise. In the case of the
exponential distribution that we selected, this fraction is $1 -
exp(-3\sigma/\langle A \rangle)$, which is $\sim 26 \%$ if
$\langle A \rangle / \sigma = 10$.

The shape of each pulse is modeled as a sudden level jump of
amplitude $A$, followed by an exponential decay with a time
constant $\tau$ :

\begin{equation}
D(t)=Ae^{-t/\tau}.
\end{equation}

\noindent For 10 time constants after the CR event, the simulated
data are obtained as $S(t) = D(t) + G(t)$, whereas elsewhere the
simulated data are simply $G(t)$. In our initial tests, we do not
include sky signals in the simulation.

We simulated chunks of 10 hours of data, to obtain a sufficient
number of events: we have an average of 72$~$000 (3600) events per
chunk in case $R_1$ ($R_2$). Our full set of simulations includes
10$~$000 simulated chunks.

\begin{figure}
\centering
\includegraphics[width=8cm]{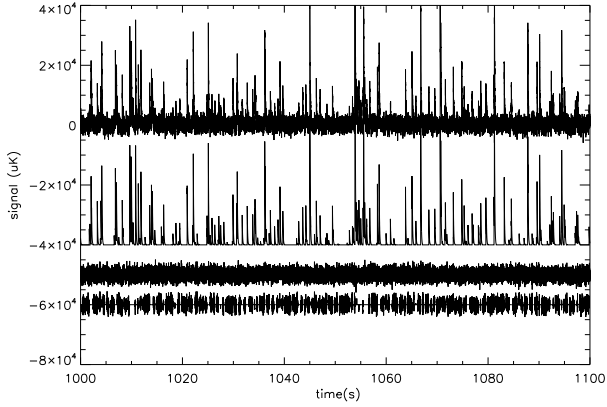}
\caption{Top line: sample subsection of simulated bolometer data
$S(t)$ (converted in $\mu K_{CMB}$) resulting from the sum of
spikes from CR events $D(t)$ (middle top line, offset -40 mK for
clarity of visualization) and bolometer noise $G(t)$ (middle
bottom line, offset -50 mK  ). In this simulation, the signal is
sampled at 200Hz, $\sigma$ = 1.4 mK, $\tau=0.07 s$, $R=2.0 Hz$,
and $\langle A \rangle =$ 10 mK (case $R_1$). Comparing the top
and middle lines, it is evident that many CR events remain hidden
in the noise and are very difficult to identify. In the bottom
line (offset -60 mK), we plot the data de-spiked following the
pixel-based procedure described in the text. The missing data have
been flagged as unusable because they are within 5 time constants
of a detected spike.
 \label{fig1}}
\end{figure}

\begin{figure}
\centering
\includegraphics[width=8cm]{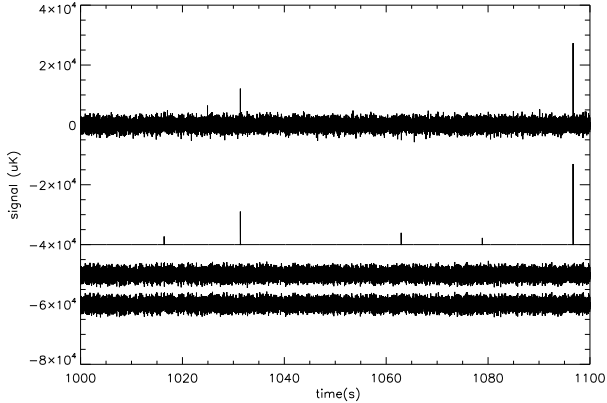}
\caption{Same as Fig. \ref{fig1}, with $\tau=0.01 s$, $R=0.1 Hz$,
and $\langle A \rangle =$ 10 mK (case $R_2$). \label{fig2}}
\end{figure}

In Figs. \ref{fig1} and \ref{fig2}, we show sample subsections of
simulated data. Large CR spikes are evident, but smaller ones
remain hidden in the noise.

\section{Despiking the time-ordered data}

Evident CR events (with amplitudes larger than say 4$\sigma$) can
be easily identified and removed. To remove smaller CR spikes, one
can take advantage of the peculiar shape of these signals. The
exact shape depends on the bolometer configuration (single or
double time constant, or even more complex response) and on the
details of the front-end electronics, amplifying the nV-level
bolometer signals. In our simulations, we used a simple
exponential decay, but it would not be difficult to extend our
analysis to more complex shapes, once the impulsive response of
the system is known from calibrations.

In the simulations described in this section, we did not include
any sky signal. With a typical 200 Hz sampling rate and 100$\mu K
/ \sqrt{Hz}$ NET, the standard deviation in the bolometer noise is
on the order of 1400 $\mu K$, while the standard deviation in the
CMB anisotropy is on the order of 100 $\mu K$. It is thus very
unlikely that CMB signals can affect the performance of any
despiking procedure. Diffuse Galactic emission can reach levels on
the order of or higher than 1000 $\mu K$ only in the Galactic
plane, which is not useful for CMB studies. Point sources are also
masked in sensitive CMB searches.

A standard method for identifying events embedded in noise is to
search for correlations between the noisy time ordered data and an
event template. To estimate the correlation, we computed the
normalized convolution

\begin{equation}
C(t)= { \int_0^{10\tau}  S(t+u) D(u) du   \over \int_0^{10\tau}
D(u) du }.
\end{equation}

\noindent In the absence of noise, this correlation is positive
for all times $t$ where a CR event contaminates the signal. Noise
induces fluctuations in $C(t)$ that, however, are smaller than the
fluctuations in $S(t)$, so there is some advantage in using this
estimator. We defined a positive threshold $T$ and assumed that
all the samples with $C(t)>T$ are contaminated by CR events. To
analyze the efficiency and accuracy of this CR detection method,
we considered the contaminated samples as a function of the
threshold level $T$, and computed the fractions $f_t$ and $f_f$.
The parameter $f_t$ is the fraction of events that have $C(t)> T$
and $D(t) > D_{min}$, where $D_{min}=1 \mu K$: these are true
detections of CR events. In contrast $f_f$ is the fraction of
samples where $C(t) > T$ while $D(t) < D_{min}$: these are false
detections caused by noise mimicking CR events. For an average
amplitude of the CR events equal to 7 times the rms of the noise
($\langle A \rangle = 10 \sigma$), we found that $50\%$ (0.4$\%$)
of the samples are contaminated in case $R_1$ ($R_2$).

Owing to the noise, we found that the convolution-based method
fails to identify a number of contaminated samples, and introduces
a large number of false detections (see Table
\ref{table:fractions} for $\langle A \rangle = 10 \sigma$). To
avoid a large number of false detections, we found that one has to
raise the threshold level, although many CR events are then
missed. The situation is even worse if the ratio $\langle A
\rangle / \sigma$ is lowered.

\begin{table}\begin{center}
\begin{tabular}{|c|c|c|c|c|}
\hline
 Threshold $T$ & $f_t (R_1)$(\%)  &  $f_f (R_1)$(\%) & $f_t (R_2)$(\%)  &  $f_f (R_2)$(\%)\\
\hline \hline
 0.5$\sigma$ & 37 &  2.9    &  33 &  19  \\
 \hline
 1$\sigma$    & 27 &  0.3    &  24 & 3.8  \\
 \hline
 2$\sigma$   & 16 &  0.005  &  13 & 0.02 \\
\hline

\end{tabular}
\caption{Fraction of true detections ($f_t$) and false detections
($f_f$) versus threshold level $T$ for the convolution-based
despiking method, for the two reference cases ($R_1$ and $R_2$)
(see text). Here $\langle A \rangle = 10 \sigma$}.
\label{table:fractions}
\end{center}
\end{table}

\section{Pixel-space despiking}

In addition to the problems listed in the previous section,
removing CR pulses directly from the time-ordered data is
inadvisable, because of the risk of removing true sky signals (for
example, fast signals produced by scans over point sources, such
as planets used as calibrators).

A more effective strategy is to analyze the set of data samples
contributing to the same sky pixel: for all of these samples, the
sky signal is the same, so outliers must be caused by either
detector noise or CR events. This is true in the absence of
pointing errors. In areas with steep brightness gradients (for
example near the Galactic plane), the combination of pixelization
and pointing errors can lead to spurious glitches. We neglect
these effects here, since they are not relevant to CMB studies.

We used a simple iterative procedure to remove outliers. For a
given pixel $p$,  the average $\langle S \rangle_p$ and the
standard deviation $\sigma_p$, in all the contributing
time-ordered signals $S_k$ were computed. For sample $k$, if the
difference $S_k - \langle S \rangle_p$ was larger than
$3\sigma_p$, the sample was classified as an outlier, and removed.
Detection of outliers was therefore performed in pixel space, but
outliers were removed in the time domain. New values of both the
average and variance were computed from the remaining samples of
the set, and new outliers were identified and removed. This
procedure was repeated until the average and the variance no
longer changed (no or few new outliers are found).

In general, only a few iterations are needed to achieve
convergence. We found that after 6 iterations the efficiency of
this removal method is similar to that of the convolution-based
one, yet with this method we do not remove true sky signals.

\begin{figure}
\centering
\includegraphics[width=8cm]{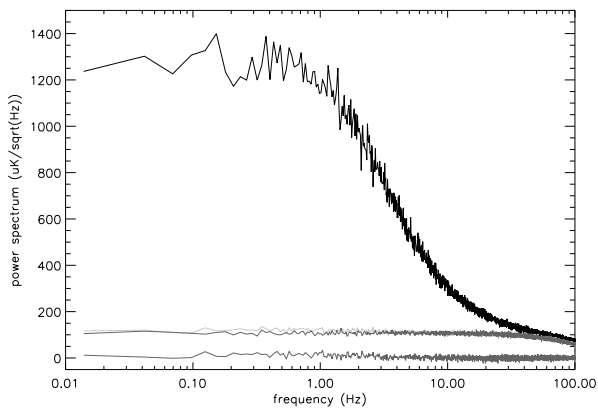}
\caption{Power spectrum of the data for case $R_1$ (top line), of
the despiked data, and of the original noise-only data (middle
lines).  The despiking procedure does not introduce spectral
features in the data, as is evident from the difference between
the despiked spectrum and the original noise-only spectrum (bottom
line). \label{fig:ps}}
\end{figure}

The despiked timelines are very similar to the noise-only
time-lines, and their spectrum (in frequency domain) closely
resembles the noise spectrum, as shown in Fig. \ref{fig:ps} for
our worst-case $R_1$.

The average in each pixel converges to a positive value, while
according to detector noise only it would be very close to zero.
This is due to CR hits of amplitude smaller than the noise: these
remain undetected and produce a positive bias. They also increase
the standard deviation in the data contributing to the same pixel.

For this reason, the 1-point distribution of temperatures in
pixels obtained after the averaging and the outlier-removal
processes will be shifted towards positive values, and be broader
than expected from instrument noise only.

\begin{figure}
\centering
\includegraphics[width=8cm]{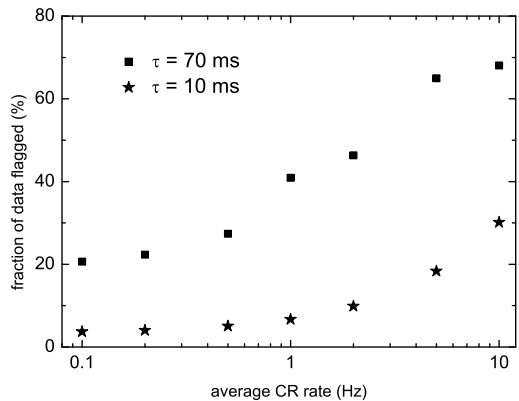}
\caption{Fraction of data flagged as contaminated by CR events,
versus average CR rate, in a bolometric experiment. A detector
noise of 100 $\mu K / \sqrt{Hz}$ and an average amplitude of CR
events of 10 mK have been assumed. Squares refer to fast detectors
(time constant of 10 ms), stars refer to slow detectors (time
constant of 70 ms). A pixel-based de-spiking algorithm iteratively
clipping all the data at more than 3$\sigma$ from the pixel
average has been used. All data within 5 time constants of a CR
event have been flagged.
 \label{figflags}}
\end{figure}

After a CR event, the data are contaminated for a certain amount
of time, depending on both the spike amplitude and the time
constants of the system. Some of these data might be recovered by
removing a best-fit spike profile from the timeline. However, due
to noise, the fit will not be perfect. We prefer to be
conservative, and flag as unusable all the data after a spike, for
a period of 5 time constants. The fraction of flagged data is
plotted in Fig. \ref{figflags} for different CR rates and
bolometer time constants. The obvious effect of CR events is thus
a significant decrease in the redundancy of the maps. Only the use
of fast detectors mitigates this problem. In the following, we
focus on the second effect, \emph{i.e.} the introduction of bias
and non-Gaussian features in the maps.

We study two different cases. The first case is a targeted
observation of a small map (about $1.5^o \times 1.5^o$ in size,
with a 4$^\prime$ FWHM resolution) for a relatively short time (10
hours of integration for a single detector), such as the maps
observed by the OLIMPO experiment to measure the Sunyaev-Zeldovich
effect in well known clusters (Nati et al. \cite{Nati07}). The
second is the observation of a large (about $10^o \times 10^o$ in
size) deep survey of a clean region, with 7 days of integration.
This is similar to the blind deep survey of CMB and undetected SZ
clusters carried out by OLIMPO. We assume the same angular
resolution for this survey as well.

\begin{figure}
\centering
\includegraphics[width=5.6cm]{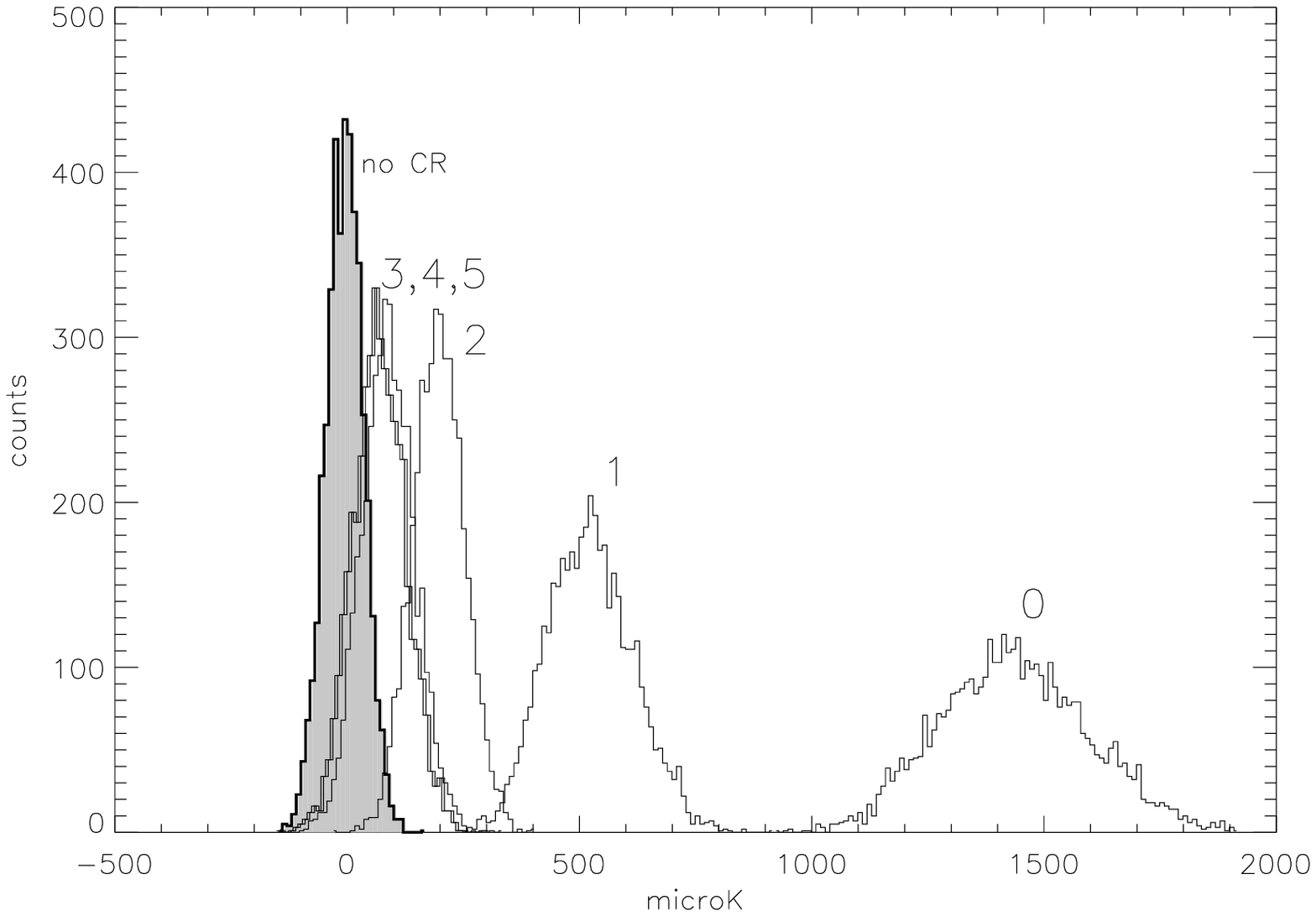}
\includegraphics[width=2.5cm]{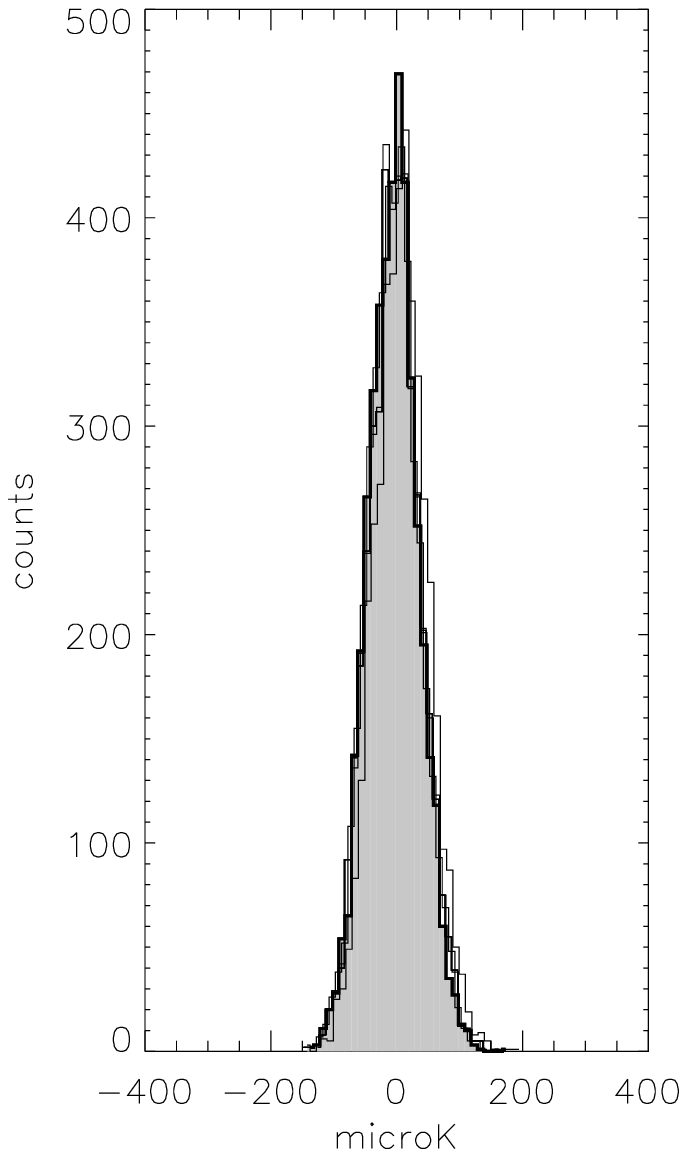}
\caption{{\bf Left}: One-point distributions of pixel temperatures
estimates $\langle S \rangle_p$, vs. despiking iteration. In this
simulation, there is no sky signal, the detector NET is 100 $\mu K
\sqrt{s}$, the time constant is 70 ms, the sampling rate is 200 Hz
($\sigma = 1.4 mK$), and the total integration time for 4320
pixels is 10 hours, in an environment producing an average
amplitude of CR events of 10 mK and an average rate of 2 Hz (case
$R_1$). The distributions are labeled with the number of outlier
removal iterations. After each iteration, the 1-point distribution
moves to the left, approaching a Gaussian distribution but
remaining shifted with respect to the distribution of detector
noise without CR events (which is the leftmost histogram, hatched
and labeled "no CR"). {\bf Right}: Here the time constant is 10
ms, in an environment producing an average amplitude of CR events
of 10 mK and an average rate of 0.1 Hz (case $R_2$): the different
distributions are hardly distinguishable.
 \label{fig3}}
\end{figure}

In Fig. \ref{fig3}, we plot the 1-point distributions versus the
iteration number in the case of observations of 4320 independent
pixels (small map case), in our two reference cases, for a total
integration time on the map of 10 hours. The results for the
moments of these distributions are reported in Table \ref{tab1}.

It is clear that the outlier removal process is effective, since
the 1-point distribution approaches a Gaussian distribution after
a few iterations. The skewness and kurtosis of the 1-point
distribution return to values consistent with zero, after the
despiking process. This means that the number of low level spikes
averaged in each pixel is large enough to produce a more Gaussian
distribution, at least at the level of sensitivity considered
here.

However, the positive bias and the increase in the variance in the
estimated pixel temperatures caused by undetected CR persist, as
is evident from Fig. \ref{fig3} and Table \ref{tab1}.

If the average rate of CR hits is not constant during the
observations of different pixels, this positive bias varies
accordingly, introducing fake structures in the measured maps. A
modulation of the CR rate can be introduced by the scanning
strategy of the instrument, which induces variations in the
absorbing material in-between the CR flux and the bolometers, thus
producing scan-synchronous systematic effects. For example, an
asymmetric satellite spinning in the solar wind could introduce
spurious dipole or higher order anisotropy (depending on the
structure of the satellite) aligned to the solar wind direction.
It would be very useful to include in the instrumental setup an
independent CR flux monitor, as close as possible to the focal
plane bolometers. The level of the spurious signal depends on the
CR composition, flux and spectrum, on the cross-section of the
bolometers, and on the structure of the instrument; this can be
quantified only by focusing on the specific case.

Even if the CR flux is perfectly steady, the increase in the
variance of the detected data for each pixel can bias the noise
estimates,  which are needed to estimate the power spectra. In the
standard analysis procedure for CMB maps one estimates the noise
from the data, assuming Gaussian noise, and then uses Gaussian
Monte Carlo simulations to assess the bias in the power spectrum
(see \emph{e.g.} Hivon et al. \cite{Hivo02}, Polenta et al.
\cite{Pole05}).

\begin{table*}[p]
\begin{center}
\begin{tabular}{|c|c|c|c|c|}
\hline Iteration & average ($\mu K$) & rms ($\mu K$) & skewness &
kurtosis \\ \hline

\hline no CR & 0.0 $\pm$ 0.7  & 39.6 $\pm$ 0.4 & -0.001 $\pm$
0.038 & 0.022 $\pm$ 0.078 \\

\hline

 0 (R$_1$) &  (1439 $\pm$ 6) &  (159 $\pm$ 4) &
(0.193 $\pm$ 0.070) & (0.06 $\pm$ 0.15) \\

1 (R$_1$) &  (529 $\pm$ 3 ) & (92.3 $\pm$ 2.1) &  (0.231 $\pm$
0.070) &  (0.11 $\pm$ 0.15)
\\

2 (R$_1$) &  (173.8 $\pm$ 1.5) &  (59.1 $\pm$ 0.8) & (0.092 $\pm$
0.042) &  (0.059 $\pm$ 0.087)\\

3 (R$_1$) &  (94.3 $\pm$ 1.1) & (56.0 $\pm$ 0.6) &  (0.020 $\pm$
0.038) &  (0.029 $\pm$ 0.076)
\\

4 (R$_1$) &  (81.4 $\pm$ 1.1) &  (57.4 $\pm$ 0.6) &  (0.014 $\pm$
0.038) &  (0.029 $\pm$ 0.076)
\\

5 (R$_1$) &  (79.3 $\pm$ 1.1) &  (57.9 $\pm$ 0.6) &  (0.012 $\pm$
0.038) &  (0.029 $\pm$ 0.077 )
\\

\hline

0 (R$_2$) &  (12.5 $\pm$ 0.7) &  (41.9 $\pm$ 0.5) &  (0.069 $\pm$
0.039) &  (0.089 $\pm$ 0.087)\\

1 (R$_2$) & (1.1 $\pm$ 0.7 ) & (40.6 $\pm$ 0.4) &  (0.001 $\pm$
0.037) & (0.029 $\pm$ 0.078) \\

2 (R$_2$) &  (0.8 $\pm$ 0.7) &  (40.9 $\pm$ 0.4) &  (0.001 $\pm$
0.037) &  (0.029 $\pm$ 0.077)\\

3 (R$_2$) &  (0.8 $\pm$ 0.7) &  (41.0 $\pm$ 0.5) &  (0.001 $\pm$
0.037) &  (0.029 $\pm$ 0.078)
\\

4 (R$_2$) &  (0.8 $\pm$ 0.7) &  (41.0 $\pm$ 0.4) &  (0.001 $\pm$
0.037) &  (0.029 $\pm$ 0.078)
\\

5 (R$_2$) & (0.8 $\pm$ 0.7) &  (41.0 $\pm$ 0.4) &  (0.001 $\pm$
0.037) &  (0.029 $\pm$ 0.078)
\\

\hline

\end{tabular}

\end{center}
\begin{center}
\caption{\small Parameters of the 1-point distribution of pixel
temperature estimates, versus iteration of the pixel-based
despiking, for cases $R_1$  and $R_2$, (observations of small
maps, same conditions as in Fig. \ref{fig3}). Iteration 0 refers
to the values without despiking. The first line reports the
noise-only case (no CRs). The errors describe the dispersion in
the results of 3000 simulations. \label{tab1}}
\end{center}
\end{table*}

In principle, this can bias the noise estimates, and, at a lower
level, might affect the non-Gaussianity parameters relevant to
cosmology.

These effects are studied in the following paragraphs, where we
focus on observations of the larger survey, which is most useful
to minimizing cosmic variance and measuring the power spectrum of
the CMB and non-Gaussianity parameters. This $\sim 100$ square
degrees map contains $N = 86~070$ $1.7^\prime$ pixels. A total
integration time of 1 week is assumed.

We define a data timeline contaminated by spikes for the cases
$R_1$ and $R_2$ described above (70 and 10 ms detector time
constant, respectively). These are, respectively, a hard case and
a mild case, from the point of view of CR contamination. We thus
perform iterative despiking as described above, and analyze the
resulting large maps.

\section{Tests of Gaussianity}

Primordial CMB anisotropies are known to have Gaussian
distributions to leading order (see \emph{e.g.} Komatsu et al.
\cite{wmap_ng}, De Troia et al. \cite{Detr03}, Natoli et al.,
\cite{Nato09} ). Small deviations from Gaussianity have however
been predicted and, if observed, may be used to constrain
inflationary models (see Bartolo et al. \cite{bartolo_ng} for a
review).

Measurements of this non-Gaussian component are however hampered
by several sources of instrumental or non instrumental
(\emph{e.g.}, foreground contamination) systematic effects, which
may induce spurious non-Gaussian signatures in the data.

In particular, undetected cosmic-ray hits add a non-Guassian
contribution to detector noise, which is usually assumed to be
otherwise normally distributed. Furthermore, the despiking
procedures described above are based on clipping and may also
alter the statistics of the dataset.

We first perform a Kolmogorov-Smirnov (KS) test on the pixel
values for (1) five iteration despiking for the $R_1$ and $R_2$
and (2) no CR contamination, the latter to be used as the
reference purely Gaussian dataset. The KS test (see Press et al.
\cite{nr}) is a standard test to quantify the probability that a
sample can be ascribed to a given distribution. It can also be
used to reject the null hypothesis that two different samples are
drawn from the same distribution. For each map, we compute the KS
measure
\[
D_\mathrm{obs} = \max_t | S_N(t) - F(t) |,
\]
where $t$ is the map temperature value, $F$ the cumulative of the
Gaussian distribution function, and $S_N$ the empirical
distribution function of a simulated noise map. We also define
$Z_\mathrm{obs} = D_\mathrm{obs}\sqrt{N}$. The KS test provides a
probability $P(Z > Z_\mathrm{obs})$ for a given map.
\begin{figure}
\centering
\includegraphics[width=8cm]{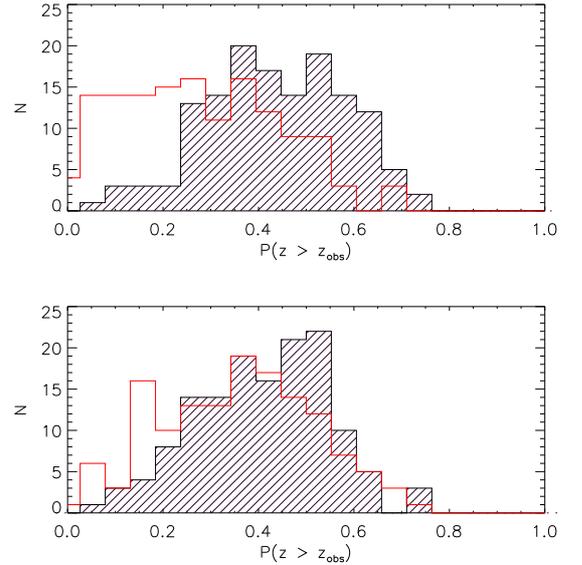}
\caption{Histogrammed probabilities of the KS test, for the $R_1$
(top) and $R_2$ (bottom) cases. The hatched histograms refer to
the reference Gaussian sets. \label{KS_prob}}
\end{figure}
In Fig.~\ref{KS_prob}, we plot the KS probabilities obtained for
the 140 Monte Carlo maps as histograms for the $R_1$ (top) and
$R_2$ cases, along with their respective Gaussian (no CR but
rescaled effective variance \footnote{The reference Gaussian
dataset is different in the two cases because the noise
 realizations are different and their variances are
corrected to account for the increase in noise caused by despiking;
 see the next section for further details.})
reference set (hatched). The $R_1$ case exhibits a clear deviation
from its reference Gaussian counterpart, while in the case of
$R_2$ this deviation is not at all evident. Even for $R_1$, any
single map has (roughly) a 50\% probability of being flagged as
non-Gaussian by our KS analysis. The same considerations apply to
the observed distribution of the $Z$ coefficients, shown in
Fig~\ref{KS_coeff}.
\begin{figure}
\centering
\includegraphics[width=8cm]{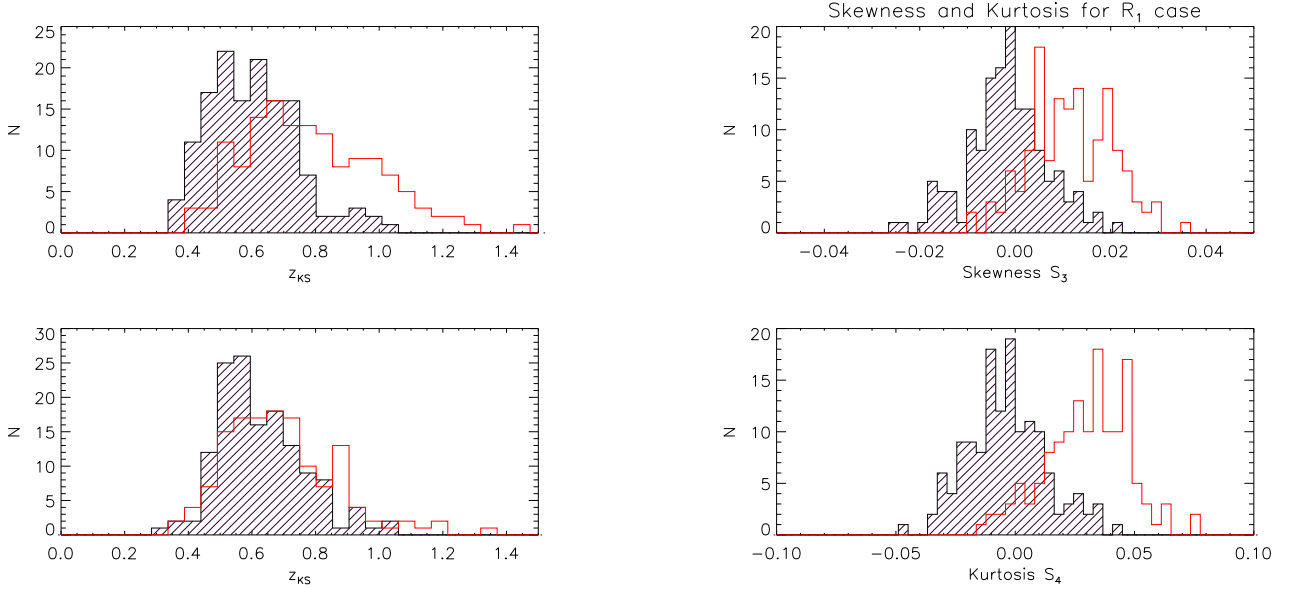}
\caption{Same as Fig.~\ref{KS_prob} but for the $Z_\mathrm{obs}$
KS coefficients. \label{KS_coeff}}
\end{figure}

To quantify our detection of non-Gaussianity, we define two simple
NG estimators: the normalized skewness ($S_3$) and kurtosis
($S_4$) of each map.
\begin{figure}
\centering
\includegraphics[width=8cm]{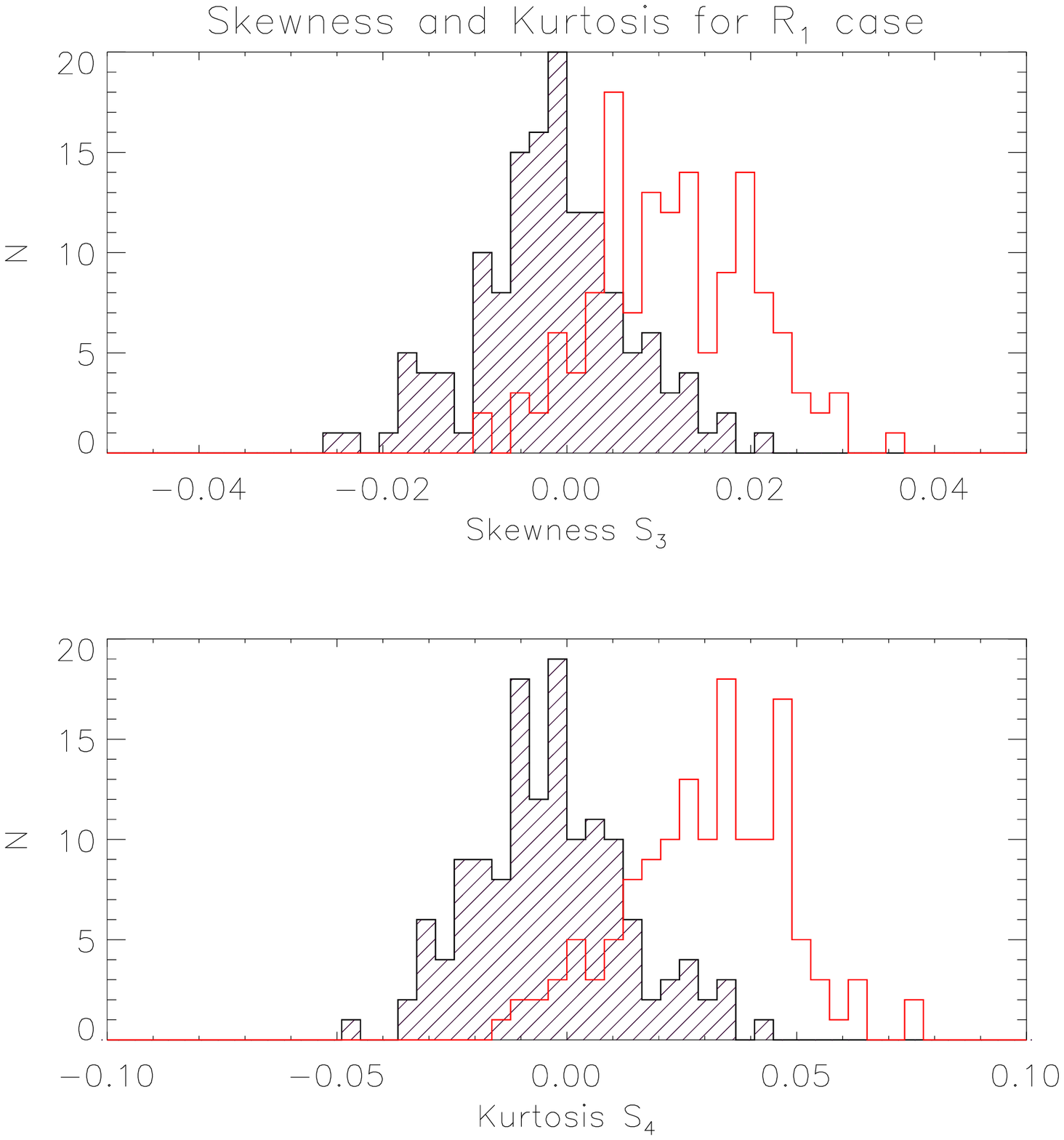}
\includegraphics[width=8cm]{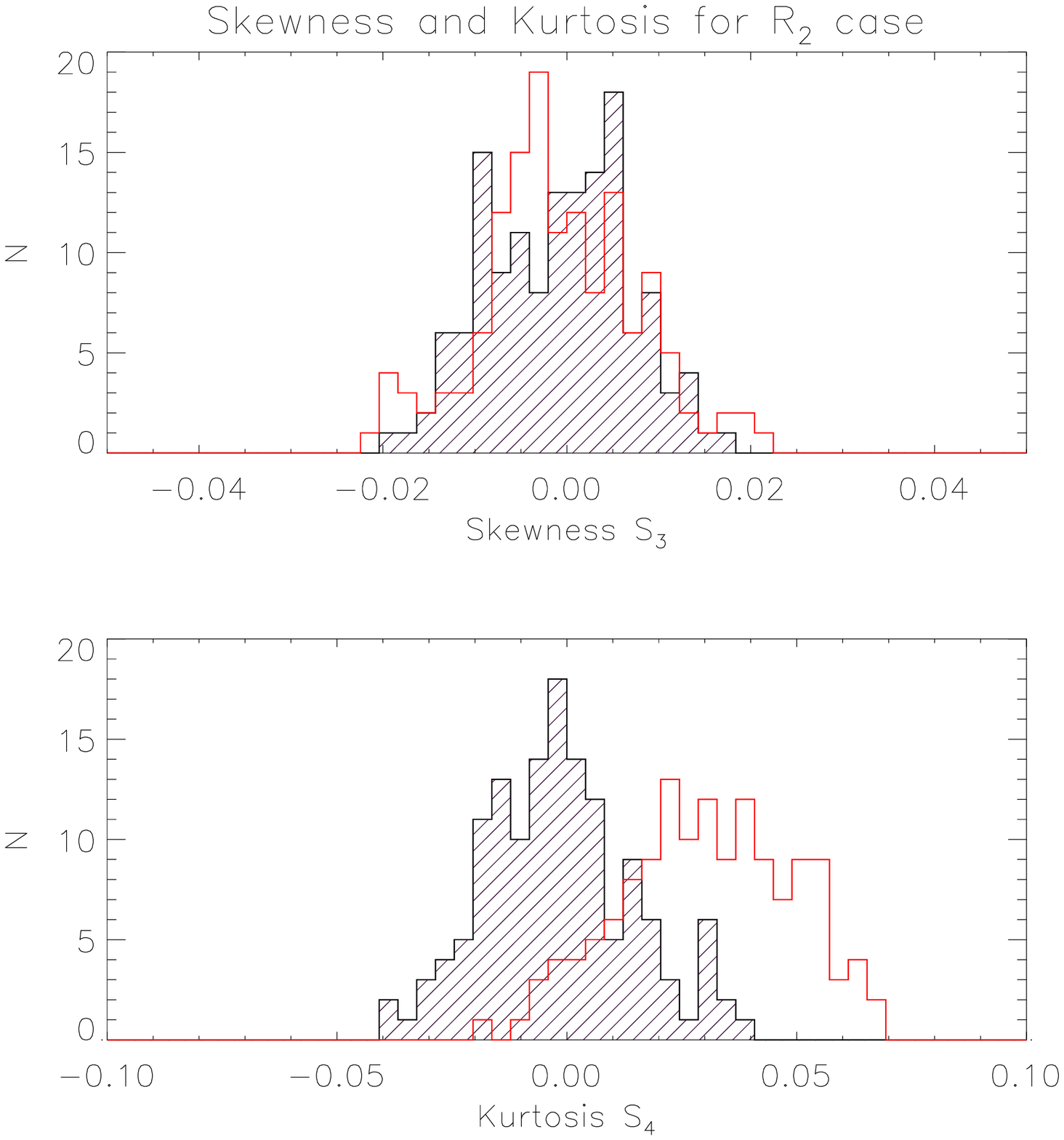}
\caption{Empirical observed probabilities for skewness ($S_3$) and
kurtosis ($S_4$), sampled from 140 Monte Carlo noise maps (red).
Shown are again the $R_1$ (top) and $R_2$ cases. The reference
Gaussian sets are shown hatched. When comparing to the values
reported in Table \ref{tab1} it should be noted that the
histograms here are computed over a considerably larger number of
pixels. \label{skew_curt}}
\end{figure}
The corresponding histograms are shown in Fig.~\ref{skew_curt},
again for the cases $R_1$, $R_2$, and their respective reference
Gaussian sets (hatched). We note how the $R_2$ case also exhibits
deviation from Gaussianity in the case of the kurtosis. The
empirical (histogrammed) distribution functions can themselves be
subject to a KS test against their Gaussian reference sets: in
this case, one simply measures the distance between the empirical
cumulative distribution functions. This test rejects at high
significance ($> 99.99 \%$) the null hypothesis that the
distribution of $S_3$ and $S_4$ is Gaussian for the $R_1$ case.
For $R_2$, the Gaussianity of $S_4$ is also rejected with similar
confidence, but the null hypothesis is accepted ($P > 35\%$) for
the $S_3$ case.

In conclusion, the non-Gaussianity level induced by CR
contamination in the large noise maps is weak and cannot be
detected with high statistical significance even in the $R_1$
case, while the $R_2$ maps are, taken one at a time,
indistinguishable from Gaussian maps, at least for the tests
employed here. However, the situation changes when a \textit{set}
of maps is considered: our Monte Carlo analysis shows that, when a
moderate number (140 in our case) are analyzed jointly,
non-Gaussian signatures are detected even for $R_2$. As a
consequence, an experiment that uses a similar detector but
gathers significantly more data than considered here, may be prone
to CR-induced non-Gaussianity, especially when high precision
measurements are sought. This is especially critical for high
sensitivity measurements of the $f_{NL}$ parameter that quantifies
the level of non-Gaussianity in the primordial perturbations
(Bartolo et al. \cite{bartolo_ng}). In particular, Planck (The
Planck Collaboration, \cite{bluebook}) is expected to tighten
existing constraints on $f_{NL}$ by roughly an order of magnitude.
We leave a detailed study of the level of contamination by
residual CR on $f_{NL}$ estimates to a future paper.

\section{Biasing of power spectra}

The angular power spectrum of the CMB is a most valuable
cosmological observable, which can be used to extract information
about the underlying physical model. Thus, it is critical to
assess the effect of contamination from residual, undetected
spikes.

We thus estimate the angular power spectrum using cROMAster, a
pseudo-$C_{\ell}$ estimator based on MASTER (Hivon et al.
\cite{Hivo02}), which was originally developed for (and applied
to) BOOMERanG-B03 ( Jones et al. \cite{Jone06}, Piacentini et al.
\cite{Piac06}, Montroy et al. \cite{Mont06}, Masi et al.
\cite{Masi06}) and later improved for PLANCK data analysis.
cROMAster can estimate both auto- and cross-power spectra, where
the former are more efficient but require noise removal and the
latter are less efficient but are naturally unbiased (Polenta et
al. \cite{Pole05}), therefore residual bias is not expected in the
pure cross-spectrum case. We verified this assertion using
simulations. The case of auto-spectra is more delicate since they
require a companion Monte Carlo dataset to estimate and remove
residual noise.

\begin{figure}
\centering
\includegraphics[width=8cm]{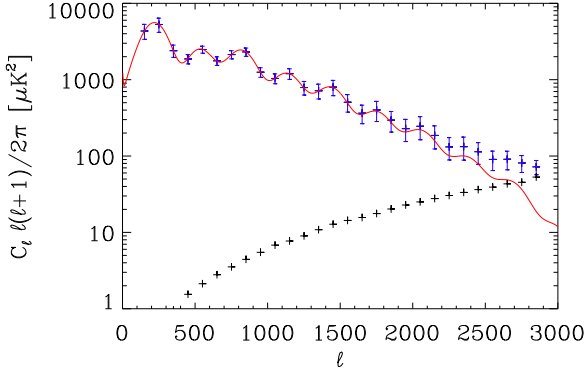}
\caption{Simulated measurement of the power spectrum of CMB
anisotropy in a bolometric experiment affected by CR hits as in
case $R_1$. The noise estimate neglects the presence of CR spikes.
As a result, the spectrum is heavily biased (black crosses show
this residual).
 \label{fig:bias}}
\end{figure}

To verify the effect of the spikes, we firstly generated a Monte
Carlo dataset that ignores the presence of residual spikes
themselves, and only relies on the nominal detector noise level.
When using this dataset as a noise estimate, the resulting power
spectrum is heavily biased, as shown in Fig.~\ref{fig:bias}.

\begin{figure}
\centering
\includegraphics[width=8cm]{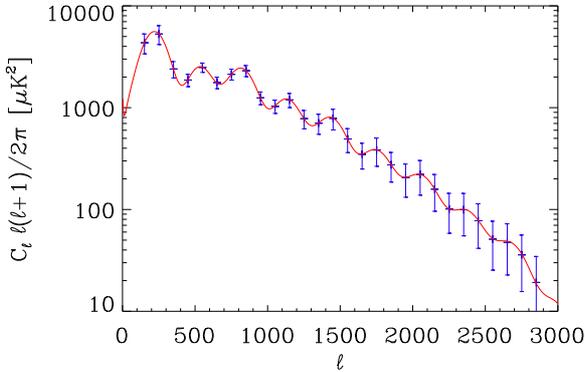}
\caption{Same as Fig. \ref{fig:bias}. Here the noise estimate
accounts for CR spikes, assuming that their distribution is known
a-priori. As a result, the spectrum is not biased.
 \label{fig:nobias}}
\end{figure}

Thus taking into account the effect of spikes in the Monte Carlo
dataset is important. This can be done in two ways: the first and
in principle more precise method is to process the Monte Carlo
timelines by adding simulated spike signal with the same
properties of the spikes seen in real data. When this is
performed, a bias free auto-spectrum can be obtained as shown in
Fig.~\ref{fig:nobias}.

However, this procedure relies on the knowledge of the underlying
spike's rate and distribution. The spike characteristics and
statistics can in principle be extrapolated from the data
themselves, relying on the observed events. This procedure,
however, significantly complicates the noise estimation and Monte
Carlo generation pipeline of the experiment. Moreover, this
approach is prone to biases, since the estimated statistics are
derived from the most energetic CR events, which are detected
above the noise, while the low energy ones, which contaminate the
data, are not detected: their distribution can only be estimated
by means of analytic extrapolation of the high energy
distribution. The reliability is thus unknown. The variance
produced by the extrapolated low-energy part of the distribution
can be used to guess the importance of the problem. The only way
to obtain robust estimates is to perform detailed and extensive
Monte Carlo simulations (using packages such as GEANT-4, see
Agostinelli et al. \cite{Agosti}) of the interaction of the full
instrument with the cosmic environment. If the high-energy part of
the simulated spectrum fits the data, the low energy part of the
simulated spectrum can be used to estimate the noise bias from
undetected spikes. This procedure is certainly complex and
computationally very intensive.

\begin{figure}
\centering
\includegraphics[width=8cm]{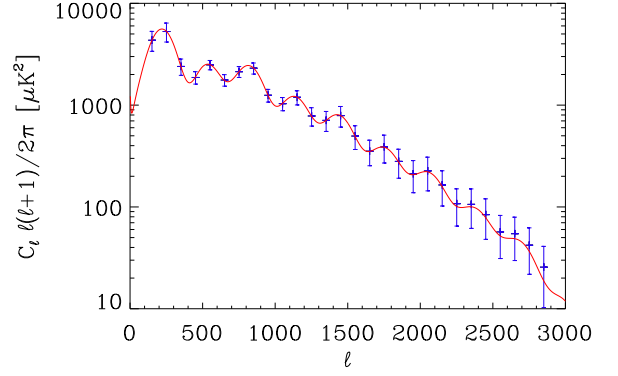}
\caption{Same as Fig. \ref{fig:bias}. Here the noise estimate has
been performed from the dataset using a standard pipeline,
\emph{i.e.} subtracting the best fit signal contribution from the
data timelines. The non-Gaussian nature of the residual spikes has
been neglected. The residual bias in the power spectrum is small,
and appears to be limited to the highest multipoles.
\label{fig:bias_small}}
\end{figure}

We thus verified wether a simpler approach would produce
sufficiently accurate results for power spectrum estimation. For
this second scheme, we estimated the noise properties from the
data that are contaminated by spikes but, in the timeline
simulation, we avoided modeling the spike contribution and
despiking procedure. In practice, we first measured the noise
properties of the despiked timeline, after subtracting the signal
contribution by means of standard iterative techniques (Ferreira
and Jaffe \cite{ferreira-jaffe-1998}), and used this dataset to
estimate the noise power spectrum. Since a timeline contaminated
by spikes exhibits some level of non-Gaussianity, the noise power
spectrum does not encode all the information about our noise
properties. However, we ignored this complication, and used the
estimated spectrum information to generate stationary Gaussian
realization of noise. Since the spike residual is probably
non-Gaussian, this procedure is not strictly correct. However, as
shown in Fig.~\ref{fig:bias_small}, the residual bias we obtain is
-visually- very small.

To quantify this statement, we performed the Hausman test (Polenta
et al.  \cite{Pole05}) on the band power we obtain from our
simulations. The Hausman test is a powerful procedure to assess
the significance of a residual noise bias in the spectrum. It uses
both cross-spectra and auto-spectra estimates. We restrict
ourselves to the $R_1$ and $R_2$ cases described above. As in
Polenta et al. \cite{Pole05}, we define three test statistics to
detect a bias in the noise estimation,
$s_{1}=\sup_{r}B_{L}(r)$, $%
s_{2}=\sup_{r}|B_{L}(r)|$, and $s_{3}=\int_{0}^{1}B_{L}^{2}(r)dr$,
where $B_{L}(r)$ is a random process defined as
\begin{equation}
B_{L}(r)=\frac{1}{\sqrt{L}}\sum_{\ell =1}^{[Lr]}H_{\ell } , r\in
\lbrack 0,1] ,
\end{equation}%
where $[.]$ denotes integer part, $L$ is the maximum multipole
used in the harmonic analysis, and $H_{\ell}$ is the difference of
the cross and auto (or in general unbiased versus biased) power
spectrum estimators normalized by its variance. It can be shown
that as $L \rightarrow \infty$, $B_L (r)$ converges to a Brownian
motion process, whose properties are widely studied and
well-known, and therefore can be used to test the null bias
hypothesis. In this case, rather than relying on the asymptotical
distribution, we used Monte Carlo simulations to draw the
empirical distributions of the test statistics. Our results for
the confidence of bias detection with 68\%, 95\%, and 99\%
probability are shown in Table \ref{table:hausman}.

\begin{table}\begin{center}
\begin{tabular}{|c|c|c|c|}
\hline
 &  68\%     &      95\%   &      99\%\\
\hline \hline
 $s_1$ & 100\%  &   96 \%  &   87\% \\
\hline $s_2$   &   99 \%   &  93\%    &  70\% \\ \hline $s_3$ &
96\% & 69\%  &   39\% \\

\hline \hline

$s_1$ & 59 \% & 19 \% & 10 \% \\ \hline $s_2$ & 48 \% & 15 \% & 7
\%
\\ \hline $s_3$ & 38 \% &  8\% & 1.4\% \\ \hline

\end{tabular}
\caption{Confidence level for bias detection with the Hausman test
in cases $R_1$ (top three lines) and $R_2$ (bottom three lines),
with a naively simulated noise estimation pipeline (see text). The
parameters $s_1$, $s_2$, and $s_3$ represent three different bias
estimators defined in Polenta et al. \cite{Pole05} (see text).}
\label{table:hausman}
\end{center}
\end{table}

We note that, despite being small, the bias can always be detected
at a high statistical significance. This is a strong indication
that the noise estimation pipeline for an high precision
experiment must account for the presence of spikes in a more
accurate way than the simple procedure set forth above, at least
if auto spectra are desired.

While we do not explicitly simulate an experiment capable of
measuring polarization spectra, it is clear that the same
conclusions hold, {\sl a fortiori}, when linear polarization can
be measured. On the other hand, a spectral pipeline that relies
only on cross-spectra is very robust to the effect of spikes.

Hence, cross-spectra, while being less efficient than
auto-spectra, are certainly more adequate for an experiment
contaminated by cosmic rays, at least as long as correlated events
between different detectors are excluded.

\section{Conclusions}

Bolometric observations carried out from space are affected by
cosmic rays. The data must be despiked to be used efficiently and
avoid serious biases in the results. Pixel-based outliers removal
works well: the power spectra of the despiked timelines closely
resemble the original Gaussian noise used in the simulation.
However, low-level events remain hidden in the noise, resulting in
a positive shift of the average signal measured in each pixel, and
increasing its variance.

The maps obtained from despiked data are non-Gaussian. The level
of non-Gaussianity depends on the rate of the spikes, on the
deposited energy and on the time constant and noise of the
detectors. Using the skewness and the kurtosis of the pixel
temperatures as simple non-Gaussianity indicators, we have found
that, in the case of the 100 square degree survey, the CR-induced
non-Gaussianity is unlikely to be detected if spider-web
bolometers are used, but would be marginally detectable for slow
membrane detectors. For experiments with longer integration times
and/or lower noise detectors, the CR-induced non-Gaussianity will
be significant. This will probably affect the constraints on the
primordial non-Gaussianity expected by present space-borne
missions. Detailed studies need to be performed in these cases.

Using the standard analysis pipeline on the map produced by the
despiked timeline, we have found that the residual hidden events
produce a positive bias in the angular power spectrum of the map
at high multipoles. In the specific example of the 100 square
degree survey, the expected bias is negligible for spider-web
bolometers, but can be important for slow membrane detectors.
However, since all modern experiments use bolometer arrays,
cross-spectra can be computed in place of auto-spectra. These
spectra are virtually unaffected by the CR-induced bias problem.
The only unavoidable problem however is a significant decrease in
the effective integration time and redundancy of the maps, in the
case of slow membrane detectors.

\acknowledgements

This work has been supported by Italian Space Agency contracts
"COFIS", "Planck-HFI" and "OLIMPO" and by PRIN 2006 ``Cosmologia
Millimetrica con Grandi Mosaici di Rivelatori'' of the Ministero
dell'Istruzione, dell'Universit\`{a} e della Ricerca.



\begin{thebibliography}{}


\bibitem[2003]{Agosti} Agostinelli, S., et al., Nucl. Inst. and Methods A, 2003, 506, 250

\bibitem[2004]{bartolo_ng} Bartolo, N., Komatsu, E., Matarrese, S., \& Riotto, A.\ 2004, \physrep, 402, 103

\bibitem[2009]{Bill09} Billot N., et al., Astrophysics Detector Workshop 2008 P. Kern (ed)
EAS Publications Series, 37 (2009) 119-125


\bibitem[2009]{Carl09} Carlstrom J.E. et al., 2009, submitted to PASP, astro-ph/0907.4445v

\bibitem[1990]{Case90} A. Caserta, P. de Bernardis, S. Masi, M. Mattioli,  Nuclear
Instrumentation and Methods in Physics Research A294, 328-334
(1990)

\bibitem[2008]{Crill08} Crill B., et al., Proceedings of SPIE Volume 7010,  "Space
Telescopes and Instrumentation 2008: Optical, Infrared, and
Millimeter", Editors: Jacobus M. Oschmann, Jr.; Mattheus W. M. de
Graauw; Howard A. MacEwen, arXiv:0807.1548

\bibitem[2003]{Detr03} De Troia, G., et al., 2003, MNRAS, 343, 284-292

\bibitem[2000]{ferreira-jaffe-1998} Ferreira, P.~G., \& Jaffe, A.~H.\ 2000, \mnras, 312, 89

\bibitem[2009]{Hind09} Hinderks J.R., et al., 2009, ApJ, 692, 1221-1246

\bibitem[2002]{Hivo02} Hivon, E., et al., 2002, Ap.J. 567, 2-17

\bibitem[2008]{Holm08} Holmes, W. A., et al., Appl. Opt. 47, 5996-6008 (2008)

\bibitem[2003]{jones03} {Jones}, W.C., {Bhatia}, R.S., {Bock},
J.J., {Lange}, A.E., 2003, SPIE, 4855, 227, astro-ph/0209132

\bibitem[2006]{Jone06} Jones, W.C., et al., 2006, ApJ, 647, 823-832

\bibitem[1984]{Kays84} Kaiser N. \& Stebbins A., 1984, Nature, 310, 391.

\bibitem[2010]{Koma10} Komatsu E., et al., 2010, Submitted to Astrophysical Journal Supplement
Series, astro-ph/1001.4538

\bibitem[2006]{Kuo06} Kuo C.L., 2006, Nuclear Instruments and Methods in Physics Research A, 559, 608-610

\bibitem[2003]{wmap_ng} Komatsu, E., et al., 2003, ApJS, 148, 119

\bibitem[2007]{Maci07} Macias-Perez, J.F., et al., 2007, Astronomy and Astrophysics,  467, 1313-1344

\bibitem[2008]{Mars08} Marsden G., et al., 2008, SPIE Conference
Proceedings, arXiv:0805.4420

\bibitem[2006]{Masi06}  Masi, S., et al., 2006, Astronomy and Astrophysics,  458 , 687-716,  astro-ph/0507509

\bibitem[1997] {mausk97} {Mauskopf}, P.D., Bock, J.J.,  Del Castillo, H., Holzapfel, W.L., Lange, A.E. 1997, Applied Optics, 36, 4

\bibitem[2006]{Mont06} Montroy T.E., et al., 2006, Ap.J., 647, 813-822

\bibitem[2007]{Nati07} Nati, F., et al., New Astronomy Reviews, 51 (2007) 385-389

\bibitem[2009]{Nato09} Natoli, P., et al., 2009, submitted to MNRAS, astro-ph/0905.4301

\bibitem[2009]{Nolt09} Nolta M., et al., 2009, ApJS, 180, 296-305

\bibitem[2004]{Oxle04} Oxley P., et al., 2004, Proc. SPIE Int. Soc. Opt. Eng., 5543,
320-331, 2004

\bibitem[2008]{Pasc08} Pascale E., 2008, et al., ApJ, 681, 400-414

\bibitem[2006]{Piac06} Piacentini F., et al., 2006, ApJ, 647, 833-839

\bibitem[2005]{Pole05} Polenta G., et al., 2005, JCAP 0511, 001

\bibitem[1992]{nr} Press, W. H., Flannery, B. P., Teukolsky, S. A. \& Vetterling, W.
T., {\it Numerical Recipes in FORTRAN, The Art of Scientific
Computing}, $2^{nd}$ Edition Cambridge University Press,
Cambridge, 1992

\bibitem[2007]{Samt07} Samtleben D., et al., 2007, Nuovo Cimento, 122B, 1353-1358

\bibitem[2009]{Saye09} Sayers J., et al., 2009, ApJ, 690, 1597-1620

\bibitem[2008]{Schu08} Schultz, B., et al., in Millimeter and Submillimeter Detectors and
Instrumentation for Astronomy IV. Edited by Duncan, William D.;
Holland, Wayne S.; Withington, Stafford; Zmuidzinas, Jonas.
Proceedings of the SPIE, Volume 7020, pp. 702022-702022-9 (2008)

\bibitem[2009]{Siri09} Siringo G., et al., A\&A, 497, 945-962

\bibitem[2008]{Swet08} Swetz D.S., et al., 2008, Proc. SPIE 7020, 702008

\bibitem[2006]{bluebook} The PLANCK Collaboration, 2006, astro-ph/0604069

\bibitem[2000]{Verd00} Verde, L., Wang, L., Heavens, A. F., \& Kamionkowski, M., 2000,
MNRAS, 313, 141

\bibitem[2008]{Wils08} Wilson G.W., et al., 2008, MNRAS, 385, 2225-2238

\bibitem[2006]{Yoon06} Yoon K. W., et al., 2006, in Millimeter and Submillimeter Detectors and Instrumentation for Astronomy III, Proceedings of SPIE, 6275, astro-ph/0606278


\end{thebibliography}
\end{document}